\documentclass[12pt]{article}

\usepackage{epsfig}

\def\bphi{\mbox{\boldmath $\phi$}}

\def\balpha{\mbox{\boldmath $\alpha$}}

\def\btau{\mbox{\boldmath $\tau$}}
\def\bnu{\mbox{\boldmath $\nu$}}

\begin{document}

\begin{titlepage}

\baselineskip 24pt

\begin{center}

{\Large {\bf The Framed Standard Model (II) - A first Test against Experiment*}}

\vspace{.5cm}

\baselineskip 14pt

{\large CHAN Hong-Mo}\\
h.m.chan\,@\,stfc.ac.uk \\
{\it Rutherford Appleton Laboratory,\\
  Chilton, Didcot, Oxon, OX11 0QX, United Kingdom}\\
\vspace{.2cm}
{\large TSOU Sheung Tsun}\\
tsou\,@\,maths.ox.ac.uk\\
{\it Mathematical Institute, University of Oxford,\\
Radcliffe Observatory Quarter, Woodstock Road, \\
Oxford, OX2 6GG, United Kingdom}

\end{center}
\vspace{.3cm}

\begin{abstract}

Apart from the qualitative features described in \cite{chm}, the renormalization
group equation derived for the rotation of the fermion mass matrices are 
amenable to quantitative study.  The equation depends on a coupling and a
fudge factor and, on integration, on 3 integration constants.  Its application 
to data analysis, however, requires the input from experiment of the heaviest
generation masses $m_t, m_b, m_\tau, m_{\nu_3}$ all of which are
known, except for
$m_{\nu_3}$.  Together then with the theta-angle in the QCD action,
there are in
all 7 real unknown parameters.  Determining these 7 parameters by fitting to
the experimental values of the masses $m_c, m_\mu, m_e$, the CKM elements
$|V_{us}|, |V_{ub}|$, and the neutrino oscillation angle
$\sin^2\theta_{13}$, one
can then calculate and compare with experiment the following 12 other
quantities $m_s,
 m_u/m_d, |V_{ud}|, |V_{cs}|, |V_{tb}|, |V_{cd}|, |V_{cb}|, |V_{ts}|, 
|V_{td}|, J, \sin^2 2\theta_{12}, \sin^2 2\theta_{23}$, and the results all
agree reasonably well with
data, often to within the stringent experimental error now achieved.  Counting
the predictions not yet measured by experiment, this means that 17 independent
parameters of the standard model are now replaced by 7 in the FSM.

\end{abstract}

* Invited talk given by TST at the Conference on 60 Years of
Yang-Mills Gauge Theories, IAS, Singapore, 25-28 May 2015.

\end{titlepage}

\clearpage

In this talk I shall endeavour to quantify the conclusions derived from the
framed standard model as described in the previous talk \cite{chm}, 
to put actual 
numbers on what were estimates or
inequalities, and to show that indeed the theory is capable of giving
a reasonable overall fit to all the data on quark and lepton masses
and mixing.

We shall start with a very brief summary of the framed standard model
(FSM)
\cite{efgt, dfsm, tfsm},
pointing out only those of its salient features which we shall refer to.  In
FSM, frame vectors form part of the geometry of gauge theory, and by
promoting these into fields which we call framons, we have built into
our system
scalar fields which can play the role of the Higgs fields.  Moreover
they entail a doubling of the gauge symmetry
$$SU(3) \times SU(2) \times U(1) \times
  \widetilde{SU(3)} \times \widetilde{SU(2)} \times \widetilde{U(1)}.$$

This results in a tree-level mass matrix of rank 1 \cite{fritsch,r2m2}
$$m = m_T \balpha^{\dagger}
  \balpha$$
with no mixing at tree-level.  Since the mass matrix is
scale-dependent, under renormalization this will lead to  
non-zero lower generation masses and non-zero mixing \cite{r2m2}, 
according to the
following formulae.  Denote the state vectors (in generation space) of
the $t, c, u$ quarks respectively by {\bf t}, {\bf c}, and {\bf u} (in
the absence of a strong $CP$ phase which does not affect the masses).
Then these are obtained by:
\begin{eqnarray}
{\bf t} & = & {\balpha}(\mu=m_t); \nonumber \\
{\bf c} & = & {\bf u} \times {\bf t}; \nonumber \\
{\bf u} & = & \frac{{\balpha}(\mu=m_t) \times {\balpha}(\mu=m_c)}
   {|{\balpha}(\mu=m_t) \times {\balpha}(\mu=m_c)|},
\label{Utriad}
\end{eqnarray}

Using these vectors, the lower generation masses are determined by
\begin{eqnarray}
m_t & = & m_U,  \nonumber\\
m_c & = & m_U |{\balpha}(\mu=m_c) \cdot{\bf c}|^2, \nonumber  \\
m_u & = & m_U |{\balpha}(\mu=m_u)\cdot{\bf u}|^2,
\label{hiermass}
\end{eqnarray}

When we take into account the strong $CP$ phase $\theta_{CP}$, the
state vectors become complex and are given by:
\begin{eqnarray}
\tilde{{\bf t}} & = & \balpha(\mu = m_t), \nonumber \\
\tilde{{\bf c}} & = & \cos \omega_U \btau(\mu = m_t) - \sin \omega_U
              \bnu(\mu = m_t) e^{-i \theta_{CP}/2}, \nonumber \\ 
\tilde{{\bf u}} & = & \sin \omega_U \btau(\mu = m_t) + \cos \omega_U
              \bnu(\mu = m_t) e^{-i \theta_{CP}/2},
\label{tcutilde}
\end{eqnarray}

The same procedure applies to the down quark triad.

The direction cosines between these two triads then give the CKM
mixing matrix:
 $$V_{CKM} = \left( \begin{array}{ccc}
   \tilde{\bf u} \cdot\tilde{\bf d}  &  \tilde{\bf u} \cdot
\tilde{\bf s}  &  \tilde{\bf u} \cdot \tilde{\bf b}  \\
    \tilde{\bf c} \cdot \tilde{\bf d}  &  \tilde{\bf c} \cdot
\tilde{\bf s}  &  \tilde{\bf c} \cdot \tilde{\bf b}  \\
    \tilde{\bf t} \cdot \tilde{\bf d}  &  \tilde{\bf t} \cdot
\tilde{\bf s}  &  \tilde{\bf t} \cdot \tilde{\bf b} 
          \end{array} \right)$$
with a complex phase corresponding to the Kobayashi-Maskawa phase
giving CP violation.  Thus we see explicitly how the QCD $\theta$
angle is transformed via rotation into the Kobayashi-Maskawa phase
\cite{ato2cp}.

The case of the leptons is similar, except that we are free not to
consider a CP violating phase in the PMNS matrix (as not yet known
experimentally).

With these formulae in hand our task is to confront them  with actual data, and
our main object of interest is the rotating vector $\balpha$.  By
definition it is a unit vector, and under renormalization it rotates
on the unit sphere, tracing out a trajectory say $\Gamma$.  To study
this we compute to 1-loop the relevant
Feynman diagrams below which are self-energy diagrams involving the
exchange of framons.

\begin{center}
\includegraphics[scale=0.45]{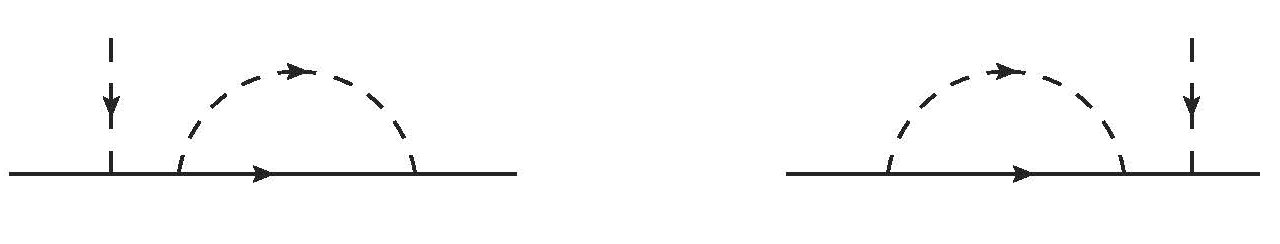}
\end{center}

\begin{center}
\includegraphics[scale=0.3]{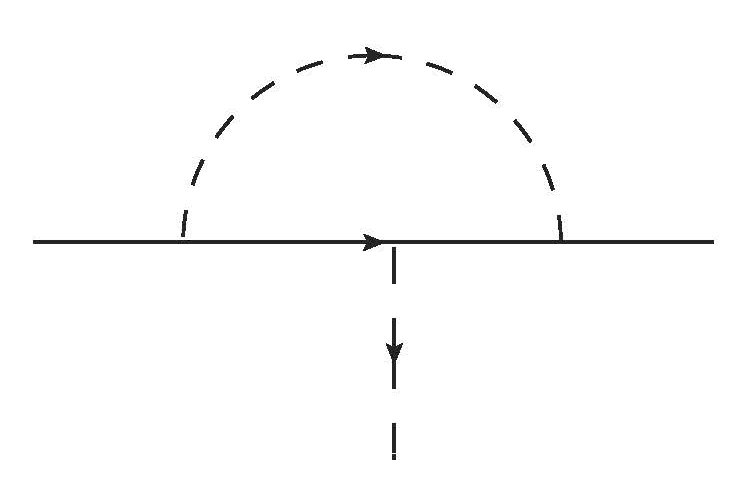}
\end{center}

\begin{center}
\includegraphics[scale=0.35]{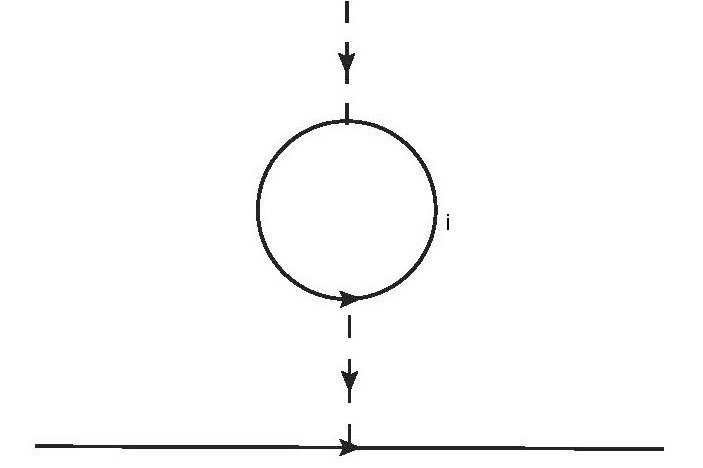}
\end{center}

Recalling that the framon potential is given by \cite{dfsm, tfsm}: 
\begin{eqnarray}
V[\balpha,\bphi,\Phi] & = & - \mu_W |\bphi|^2 + \lambda_W
(|\bphi|^2)^2  \nonumber \\
    & &      - \mu_S \sum_{\tilde{a}} |{\bphi}^{\tilde{a}}|^2
          + \lambda_S \left( \sum_{\tilde{a}}
            |{\bphi}^{\tilde{a}}|^2 \right)^2 
   +   \kappa_S \sum_{\tilde{a},\tilde{b}} 
    |{\bphi}^{\tilde{a}*}\cdot{\bphi}^{\tilde{b}}|^2 \nonumber \\
& &   + \nu_1 |\bphi|^2 \sum_{\tilde{a}} |{\bphi}^{\tilde{a}}|^2
    - \nu_2 |\bphi|^2 |\sum_{\tilde{a}} \alpha^{\tilde{a}}
    {\bphi}^{\tilde{a}}|^2
\label{potential}
\end{eqnarray}
we deduce the renormalization equation for $\balpha$.

If we write $\balpha$ in spherical polar coordinates as usual
\begin{equation}
\balpha = \left( \begin{array}{lll} \sin \theta \cos \phi \\
                                    \sin \theta \sin \phi \\
                                    \cos \theta \end{array} \right)
\end{equation}
and introduce the parameter
\begin{equation}
R=\frac{\zeta_W^2\nu_2}{2 \kappa_S \zeta_S^2},
\end{equation}
we obtain RGE for the parameters of $\balpha$ as follows:

\begin{eqnarray}
\dot{R} & = & - \frac{3 \rho_S^2}{16 \pi^2} \frac{R(1-R)(1+2R)}{D}
   \left( 4 + \frac{R}{2+R} - \frac{3 R \cos^2 \theta}{2+R} \right)
\label{Rdot} \\
\dot{\theta} & = &  - \frac{3 \rho_S^2}{32 \pi^2} 
   \frac{R \cos \theta \sin \theta}{D} 
   \left( 12 - \frac{6R^2}{2+R} - \frac{3k(1-R)(1+2R)}{2+R} \right)
\label{thetadot}
\end{eqnarray}
and
\begin{equation}
\cos \theta \tan \phi = a\  ({\rm constant})
\label{phidot}
\end{equation}
where
\begin{equation}
D = R(1+2R) - 3R\cos^2 \theta + k(1-R)(1+2R).
\end{equation}
Here a dot denotes differentiation with respect to $t=\log \mu^2$.

\begin{figure}
\centering
\includegraphics[height=17cm]{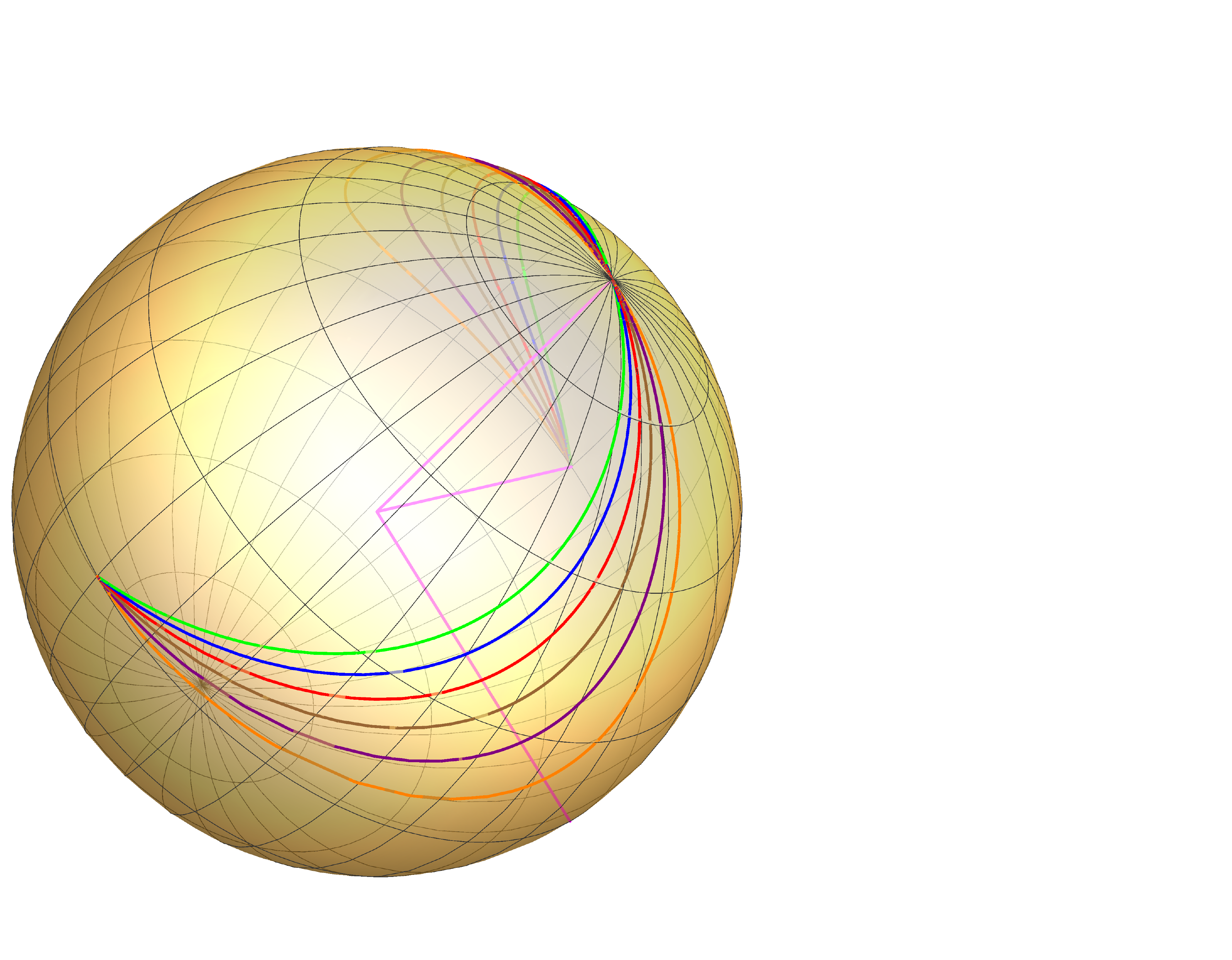}
\caption{The curve $\Gamma$ traced out by the vector $\balpha$
on the unit sphere in generation space for various values of 
the integration constant $a$, decreasing in magnitude from $a=-0.6$ in green to
$a=-0.1$ in orange.}
\label{Gammaplot}
\end{figure}

The trajectory $\Gamma$ traced out by $\balpha$ on the sphere as we
vary the scale $\mu$ depends on two functions (of scale) which we may
call the shape function and the speed function.  The shape function
is a consequence of symmetry and
depends only on one real parameter $a$ (\ref{phidot}); so it is
simple to deal with, and has been discussed in some detail in
\cite{chm} (see Figure \ref{Gammaplot}).  The speed function, 
on the other hand, is much less
precisely predicted, since first the RGE is only to 1 loop, and secondly it
depends on 3 parameters $\rho_S$ and two integration constants.  Even
more seriously it depends on some unknown and perhaps uncalculable effects
represented by the function $k(\mu)$.  Clearly one can do little
phenomenologically with an unknown function.  With some justification
we replace it by a constant $k$.  

Before we actually confront our theory with the fermion masses and 
mixing data, let us
see what kind of data we are faced with.  There are three points to
note.
\begin{itemize}
\item There is a large amount of data.
\item They have vastly different percentage errors.
\item The masses range over 13 orders of magnitude.
\end{itemize}

To be concrete, let me quote the data as given by the PDG \cite{pdg} in the
summer of 2014, when we did the fit to be reported below \cite{tfsm}.

The quark masses:
\begin{eqnarray*}
m_t &=& 173.07\pm0.52\pm0.72 \ {\rm GeV}\\
m_c &=& 1.275 \pm 0.025\ {\rm GeV}\\
m_u &=& 2.3^{+0.7}_{-0.5}\ {\rm MeV}\ ({\rm at\ 2\ GeV})\\
m_b &=& 4.18 \pm 0.03\ {\rm GeV}\\
m_s &=& 0.095 \pm 0.005\ {\rm GeV}\ ({\rm at\ 2\ GeV})\\
m_d &=& 4.8 ^{+0.5}_{-0.3}\ {\rm MeV}\ ({\rm at\ 2\ GeV})\\
\end{eqnarray*} 

The lepton masses:
\begin{eqnarray*}
m_\tau &=& 1776.82\pm0.16\ {\rm MeV}\\
m_\mu &=& 105.6583715 \pm 0.0000035 {\rm MeV}\\
m_e &=& 0.510998928 \pm 0.000000011\ {\rm MeV}\\
\end{eqnarray*}
with the squared mass differences for the neutrinos:
\begin{eqnarray*}
(m^{\rm ph}_{\nu_3})^2 - (m^{\rm ph}_{\nu_2})^2& =& (2.23^{+0.12}_{-0.08})
   \times 10^{-3} \  {\rm eV}^2\\
(m^{\rm ph}_{\nu_2})^2 - (m^{\rm ph}_{\nu_1})^2 &=& (7.5 \pm 0.20) \times
   10^{-5} \ {\rm eV}^2\\
\end{eqnarray*}

The quark CKM matrix:
$$
\left( \begin{array}{lll}
|V_{ud}| & |V_{us}| & |V_{ub}|\\
|V_{cd}| & |V_{cs}| & |V_{cb}|\\
|V_{td}| & |V_{ts}| & |V_{tb}| \end{array} \right)$$
$$\left( \begin{array}{lll} 
       0.97427 \pm 0.00015 & 0.22534 \pm  0.00065 & 
0.00351^{+0.00015}_{-0.00014} \\
       0.22520 \pm 0.00065 & 0.97344\pm0.00016 & 0.0412^{+0.0011}_{-0.0005}\\
       0.00867^{+0.00029}_{-0.00031} & 0.0404^{+0.0011}_{-0.0004} &
0.999146^{+0.000021}_{-0.000046}
\end{array} \right) $$
$$|J|=\left(2.96^{+0,20}_{-0.16} \right) \times 10^{-5}$$

The neutrino oscillation angles:

\begin{eqnarray*}
\sin^2 2\theta_{13} &=& 0.095 \pm 0.010\\
\sin^2 2\theta_{12} &=& 0.857 \pm 0.024\\
\sin^2 2\theta_{23} &>& 0.95\\
\end{eqnarray*}

The last two angles are also known as the solar angle and the
atmospheric angle, respectively.  I think we could appropriately 
call the first the Daya Bay angle.

We are now ready to vary the parameters of the theory so as to
produce a trajectory $\Gamma$ for $\balpha$, which will fit best the
masses and mixing data via the formulae (\ref{hiermass}) and
(\ref{tcutilde}). 

Remembering that all the fermions lie on the same trajectory $\Gamma$, 
we fix the positions for each type ($U$ quark, $D$ quark, charged
lepton, neutrino) by inputting the heaviest generation masses, as
shown in Table \ref{heaviest}.

\begin{table}
\centering
\begin{tabular}{|l|l|l|}
\hline
& Expt (September 2014) & Input value\\
\hline
$m_t$ & $173.07\pm0.52\pm0.72$ GeV & $173.5$ GeV\\
$m_b$ & $4.18 \pm 0.03$ GeV & $4.18$ GeV \\
$m_\tau$ & $1776.82\pm0.16$ MeV & $1.777$ GeV\\
\hline
\end{tabular}
\caption{The heaviest fermion from each type as input}
\label{heaviest}
\end{table}

The neutrino masses are assumed to be generated by some see-saw
mechanism, so we
put in some assumed value of $m_{\nu_3}$ for the Dirac mass of the
heaviest neutrino to fit
the data.  The value affects only the lepton sector.

Next we shall do a parameter count in detail.  The theory has 7
adjustable parameters (\ref{potential}),
\begin{equation}
a,\rho_S,k, R_I,\theta_I,m_{\nu_3},\theta_{CP}.
\label{param7}
\end{equation}

We shall choose experimental data to fix these, using the following
criteria for the choice:
\begin{itemize}
\item that they are sufficient to determine the 7 parameters 
adequately,
\item that they have been measured in experiment to reasonable
accuracy,
\item that they are sufficiently sensitive to the values of the
parameters,
\item that they are strategically placed in $t = \ln \mu^2$ over
the interesting range,
\end{itemize}
and we end up with the following choice: 
\begin{itemize}
\item the masses $m_c, m_\mu, m_e$ 
\item the elements $|V_{us}|, |V_{ub}|$ of the CKM matrix for
quarks
\item  neutrino oscillation angle $\sin^2 2 \theta_{13}$.
\end{itemize}
Because of the special role played by the Cabibbo angle $|V_{us}|$
with respect to the geodesic curvature of the trajectory $\Gamma$, it
fixes by itself already two of the parameters above (\ref{param7}).

We now demand that, by varing the 7 parameters (\ref{param7}),  
the calculations
give us back the 6 inputted data within the desired accuracy: either within
experimental errors or within half a percent, as shown in Table
\ref{input6}.

\begin{table}
\centering
\begin{tabular}{|l|l|l|l|l}
\hline
& Expt (June 2014) & FSM Calc & Agree to \\
\hline
&&& \\
{\sl INPUT} &&&\\
$m_c$ & $1.275 \pm 0.025$ GeV & $1.275$ GeV & $< 1 \sigma$\\
$m_\mu$ & $0.10566$ GeV & $0.1054$ GeV & $0.2 \%$ \\
$m_e$ & $0.511$ MeV &$0.513$ MeV & $0.4 \%$ \\
$|V_{us}|$ & $0.22534 \pm  0.00065$ & $0.22493$ & $< 1 \sigma$ \\
$|V_{ub}|$ & $0.00351^{+0.00015}_{-0.00014}$& $0.00346$ & $< 1 \sigma$ \\
$\sin^2 2\theta_{13}$ & $0.095 \pm 0.010$ & $0.101$ &$< 1 \sigma$ \\
\hline
\end{tabular}
\caption{The input experimental values compared with calculated values}
\label{input6}
\end{table}

Note that the functional form for the trajectory for $\balpha$
 having already been prescribed by the RGE (\ref{Rdot},
 \ref{thetadot}, \ref{phidot}), it is not at all obvious 
that the 6 targeted quantities can be so fitted with the given 
7 parameters.  That it can indeed be done to the accuracy 
stipulated constitutes already quite a nontrivial test.

This test done, we can proceed to calculate the following
23 quantities of the
standard model\footnote{By the
standard model here, we mean that in which the now established fact,
that neutrinos have masses and oscillate, is incorporated.  This 
means it will have to carry the Dirac masses of the neutrinos also
as parameters.  Further, we count $\theta_{CP}$ also as a parameter
of the standard model although it is often arbitrarily put to zero.}:
\begin{itemize}
\item 8 lower generation masses
\item the absolute values of all 9 CKM elements 
\item the Jarlskog invariant $J$
\item 3 neutrino oscillation angles
\item $m_{\nu_3}$
\item $\theta_{CP}$
\end{itemize}

Of these, 17 ($=23-6$) are independent in SM
\begin{itemize}
\item 8 lower generation masses
\item 4 CKM parameters
\item 3 neutrino oscillation angles
\item $m_{\nu_3}$
\item $\theta_{CP}$
\end{itemize}
and of which 12 ($=17-5$) can be compared to experiment (the remaining
5 being not yet measured):
\begin{itemize}
\item $m_c,m_s,m_\mu,m_e, m_u/m_d$
\item 4 CKM parameters
\item 3 neutrino oscillation angles
\end{itemize}

However, we need to check 18 ($=23-5$) experimental values to ensure
that we have good accuracy, although these are not all independent in
the standard model.  For example, although the CKM matrix is unitary
and has only 4 independent parameters, ensuring that only these 4 fall within
error does not imply that the remaining elements are within error too.
These 18 quantities we checked 
are listed in Tables \ref{input6} and \ref{output12}.

\begin{table}
\centering
\begin{tabular}{|l|l|l|l|l}
\hline
& Expt (June 2014) & FSM Calc & Agree to \\
\hline
&&& \\
{\sl OUTPUT} &&&\\
$m_s$ & $0.095 \pm 0.005$ GeV & $0.169$ GeV & QCD  \\
& (at 2 GeV) &(at $m_s$) &running \\
$m_u/m_d$ & $0.38$---$0.58$ & $0.56$ &  $< 1 \sigma$ \\
$|V_{ud}|$ &$0.97427 \pm 0.00015$ & $0.97437$ & $< 1 \sigma$ \\
$|V_{cs}|$ &$0.97344\pm0.00016$ & $0.97350$ & $< 1 \sigma$ \\
$|V_{tb}|$ &$0.999146^{+0.000021}_{-0.000046}$ & $0.99907$ &$1.65
\sigma$\\
$|V_{cd}|$ &$0.22520 \pm 0.00065$ & $0.22462$ & $< 1 \sigma$ \\
$|V_{cb}|$ & $0.0412^{+0.0011}_{-0.0005}$ & $0.0429$ & $1.55 \sigma$ \\
$|V_{ts}|$ & $0.0404^{+0.0011}_{-0.0004}$ & $0.0413$ &$< 1 \sigma$\\  
$|V_{td}|$ & $0.00867^{+0.00029}_{-0.00031}$ & $0.01223$ & 41 \% \\
$|J|$ & $\left(2.96^{+0,20}_{-0.16} \right) \times 10^{-5}$ & $2.35
\times 10^{-5}$ & 20 \%  \\
$\sin^2 2\theta_{12}$ & $0.857 \pm 0.024$ & $0.841$ &  $< 1 \sigma$\\ 
$\sin^2 2\theta_{23}$ & $>0.95$ & $0.89$ & $> 6 \%$ \\
\hline
\end{tabular}
\caption{The calculated output values using inputs in Table \ref{input6}}
\label{output12}
\end{table}

We note that of the 12 output quantities shown in Table
\ref{output12}, 6 are 
within experimental error or else ($m_\mu, m_e$)
within 0.5 percent of the accurate measured values, while 2 are 
within $\sim 1.5 \sigma$.  Of the remaining 4, 1 ($m_s$) can only 
be roughly compared with experiment, because of QCD running, 
and it does so compare quite reasonably.  The
other 3: $|V_{td}|, J, \sin^2 2 \theta_{23}$, are all outside
the stringent experimental errors, but still not outrageously
so.  Besides, $|V_{td}|$ and $J$ both being small and therefore
delicate to reproduce, obtaining them with the right order of
magnitude as they are here is already no mean task.

The fit gives in addition the following values for the 5 other
standard model parameters which, not being measured, cannot 
be checked against experiment at present:
$$
\theta_{CP} = 1.78, \ \ m_u(\mu=m_u) = 0.22\ {\rm MeV}\
[{\rm or}\ m_d(\mu=m_d) = 0.39\ {\rm MeV}],$$
$$
m_{\nu_3} = 29.5\ {\rm MeV}, \ \ m_{\nu_2} = 16.8\ {\rm MeV},
\ \ m_{\nu_1} =  1.4\ {\rm MeV}.
$$

\begin{figure}
\centering
\includegraphics[height=17cm]{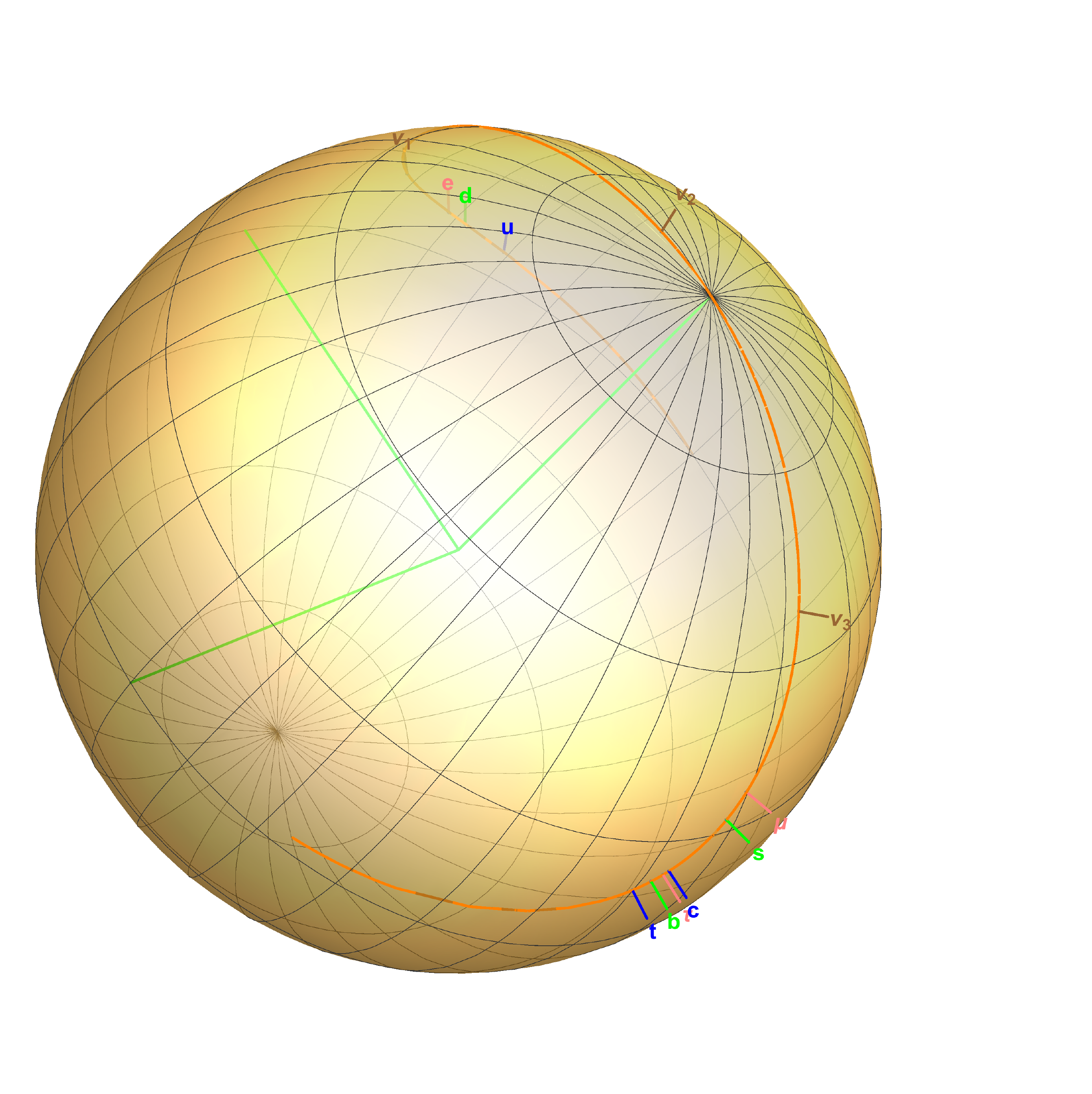}
\caption{The trajectory for $\balpha$ on the unit sphere in
generation space obtained from the parameter values obtained as described, 
showing the locations on the trajectory
where the various quarks and leptons are placed: high scales in
front.}
\label{high}
\end{figure}

\begin{figure}
\centering
\includegraphics[height=17cm]{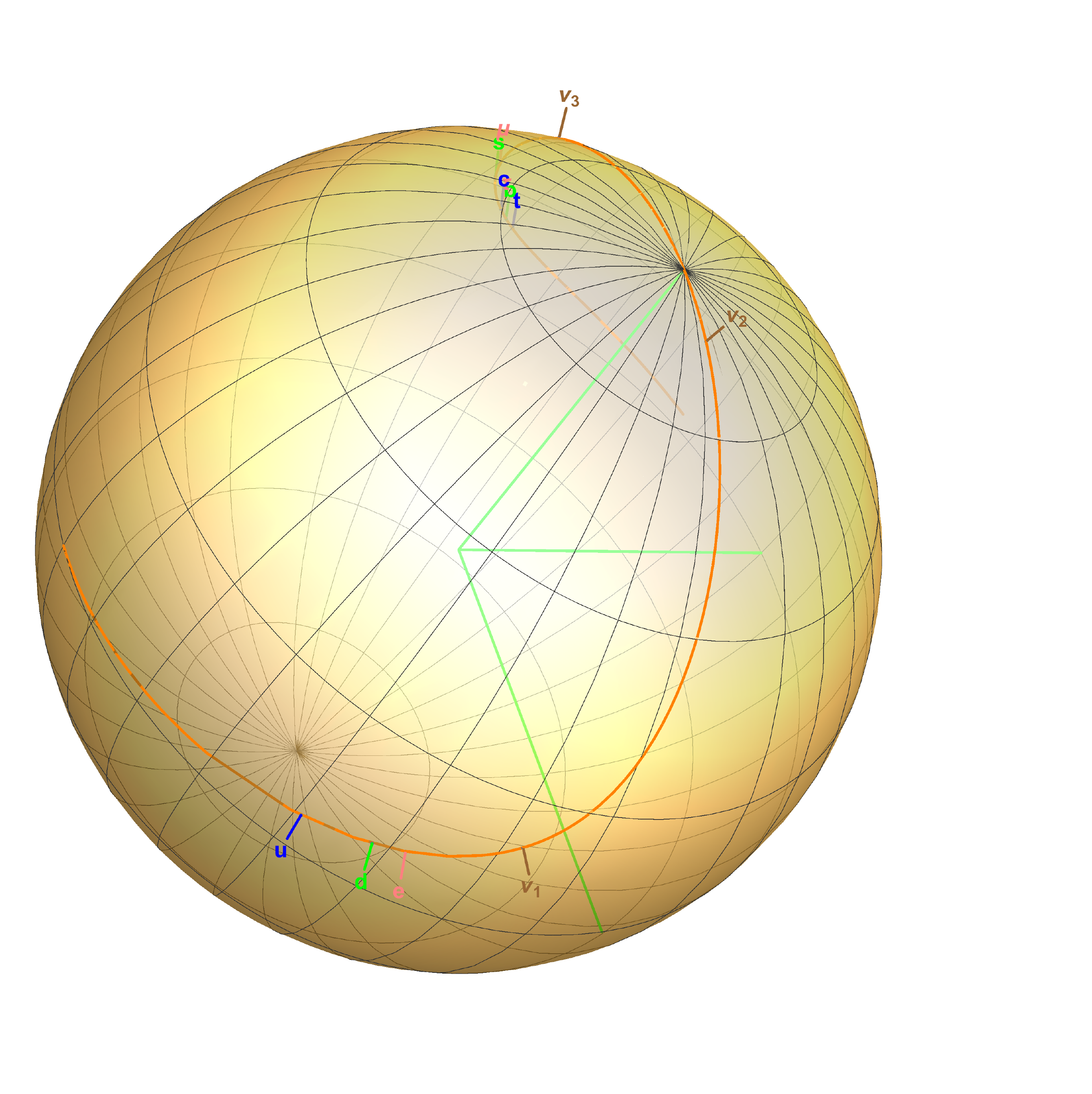}
\caption{The trajectory for $\balpha$ on the unit sphere in
generation space obtained from the parameter values obtained as described,
showing the locations on the trajectory
where the various quarks and leptons are placed: low scales in front.}
\label{low}
\end{figure}

The Figures \ref{high} and \ref{low} show the actual trajectory of
$\balpha$ corresponding to the fit above.  There are many interesting
features, in accordance with the qualitative expectations described in
\cite{chm}.  Here we would like to comment on one particular aspect.

As already mentioned in \cite{chm}, because the change in sign
of the geodesic curvature of $\Gamma$, we have, {\rm generically} as a
consequence of symmetry
and not only for this fit, that $m_u<m_d$, despite the fact that $m_t \gg m_b, 
m_c \gg m_s$.  Thus we are able to reproduce this crucial empirical
fact, without which the proton would be unstable and we ourselves
would not exist.  To understand this result a bit further, we note
that the geodesic curvature changes sign around the scale of order
MeV, where masses of the lowest generation quarks
occur.  Now according to (\ref{hiermass}) above, the mass of 
the $u$ and $d$ quarks in FSM are to be given respectively by 
solution of the equations:
\begin{equation}
|\langle {\bf u}|\balpha(\mu) \rangle|^2 = \mu, \ \ \ 
|\langle {\bf d}|\balpha(\mu) \rangle|^2 = \mu,
\label{udmass}
\end{equation}
where ${\bf u}$ the state vector of $u$ is of 
course orthogonal to ${\bf t}$ and ${\bf c}$, 
the state vectors of $t$ and $c$.  
Similarly for the triad ${\bf b}, {\bf s}, {\bf d}$.  The
masses of $u$ ($d$) being only of order MeV, this means that 
one has an approximate solution for $m_u$($m_d$) whenever the
vector $\balpha$ crosses the ${\bf t}{\bf c}$-plane (${\bf b}
{\bf s}$-plane).  Given the ordering of the masses of $t,b$ 
and that, as noted before, $m_c/m_t < m_s/m_b$, the picture is
as shown in Figure \ref{udmasspic}.  It is thus clear that in
the MeV region where the geodesic curvature has the opposite 
sign to that in the high scale region, the vector $\balpha$
must cross the ${\bf b}{\bf s}$-plane before (i.e.\ at a higher 
scale than) the ${\bf t}{\bf c}$-plane.  In other words, $m_d$ 
must be larger than $m_u$, as experiment wants.

It is interesting to note that a scale of a few MeV
(at which our geodesic curvature changes sign) occurs also, but for a
different reason, in another rotating mass scheme \cite{bj} quite
similar to ours.

\begin{figure}
\centering
\includegraphics[height=18cm]{intersectud2.pdf}
\caption{Figure illustrating the reason why $m_u < m_d$ in Table
\ref{output12}.}
\label{udmasspic}
\end{figure}

In summary, we can say that
\begin{itemize}
\item with 7 adjustable parameters 
\item can calculate 23 quantities
\item of which 18 are measurable
\begin{itemize}
\item 10 within errors
\item 2 within $0.5$ \%
\item 2 within $\sim 1.5 \sigma$ 
\item 3 within order of magnitude (or better)
\item 1 with QCD running (cannot calculate at present)
\end{itemize}
\item 17 independent in SM
\item 12 both measurable and independent
\item bonus point: $m_d>m_u$ generically
\end{itemize}

In this short talk I did not present the framed standard model
  in full, but only a small part of it: a fit to data.
We did not explore all parameter space; our purpose was
just to show that it is possible to obtain a decent (or to our
biased eyes, a good) fit.  This fit
fixes for us a number of parameters in the theory,
which we shall use to further explore consequences of FSM.

The work reported was done in collaboration with Jose Bordes.


\begin{thebibliography}{10}

\bibitem{chm} Chan Hong-Mo and Tsou Sheung Tsun,
 invited talk (by CHM) at the Conference 
  on 60 Years of Yang-Mills Gauge Field Theories, 25-28 May 2015,
  Singapore, to be published in the Proceedings, ed. Lars Brink and
  Kok Khoo Phua;  arXiv:1505.0547.

\bibitem{efgt} Chan Hong-Mo and Tsou Sheung Tsun,
  Int. J. Mod. Phys. A27 (2012) 1230002 ; arXiv:1111.3832.

\bibitem{dfsm} Michael J Baker, Jose Bordes, Chan Hong-Mo and Tsou
  Sheung Tsun, Int. J. Mod. Phys. A27 (2012) 1250087; arXiv:1111.5591.

\bibitem{tfsm} Jose Bordes, Chan Hong-Mo and Tsou Sheung Tsun, Int. J. Mod. 
   Phys. A30 (2015) 1550051; arXiv:1410.8022.

\bibitem{fritsch} H. Fritsch, Nucl. Phys. B155 (1978) 189.

\bibitem{r2m2} Michael J. Baker, Jose Bordes, Chan Hong-Mo and Tsou Sheung Tsun,
   Int. J. Mod. Phys. A26 (2011) 2087-2124; arXiv:1103.5615.

\bibitem{ato2cp} Jos\'e Bordes, Chan Hong-Mo and Tsou Sheung Tsun,
    Int. J. Mod. Phys.  A25 (2010) 5897-5911; arXiv:1002.3542
    [hep-ph].

\bibitem{pdg} PDG website: http://durpdg.dur.ac.uk/lbl/

\bibitem{bj} James Bjorken, private communication; see also the
  website\\ bjphysicsnotes.com

\end{thebibliography}
\end{document}